\PassOptionsToPackage{table}{xcolor}
\documentclass[conference]{IEEEtran}
\IEEEoverridecommandlockouts

\usepackage{cite}
\usepackage{amsmath,amssymb,amsfonts}
\usepackage{graphicx}
\usepackage{textcomp}

\usepackage{amsmath}

\usepackage{booktabs}
\usepackage[hidelinks]{hyperref}
\usepackage{enumitem,kantlipsum}

\usepackage{dirtytalk}

\usepackage{amsmath}
\usepackage{algorithm}
\usepackage{algpseudocode}
\usepackage{amsfonts}
\usepackage{amsthm}

\newtheorem{theorem}{Theorem}

\theoremstyle{definition}
\newtheorem{example}{Example}
\newtheorem{definition}{Definition}

\newcommand{\off}[1]{}
\newcommand{\verysmall}{\scriptsize}
\begin{document}

\title{An Efficient Hybrid Key Exchange Mechanism\vspace{-0.2cm}}
\author{Benjamin D. Kim, Vipindev Adat Vasudevan, Alejandro Cohen, Rafael G. L. D'Oliveira, \\ Thomas Stahlbuhk, and Muriel M\'edard

\thanks{B.~D.~Kim is with University of Illinois Urbana-Champaign, USA (e-mail: bdkim4@illinois.edu). V.~Adat Vasudevan and M.~Médard are with Massachusetts Institute of Technology, USA (email: \{vipindev, medard\}@mit.edu). A.~Cohen is with Technion, Israel (e-mail: alecohen@technion.ac.il). R.~G. L. D'Oliveira is with Clemson University, USA (e-mail: rdolive@clemson.edu). T.~Stahlbuhk is with MIT Lincoln Laboratory, USA (e-mail: thomas.stahlbuhk@ll.mit.edu).}}

\vspace{-0.4cm}
\maketitle

\begin{abstract}
We present \textsc{CHOKE}, a novel code-based hybrid key-encapsulation mechanism (KEM) designed to securely and efficiently transmit multiple session keys simultaneously. By encoding $n$ independent session keys with an individually secure linear code and encapsulating each resulting coded symbol using a separate KEM, \textsc{CHOKE} achieves computational individual security -- each key remains secure as long as at least one underlying KEM remains unbroken. Compared to traditional serial or combiner-based hybrid schemes, \textsc{CHOKE} reduces computational and communication costs by an $n$-fold factor. Furthermore, we show that the communication cost of our construction is optimal under the requirement that each KEM must be used at least once.
\end{abstract}


\vspace{-0.1cm}
\section{Introduction}
Public-key cryptography is computationally expensive and inefficient for encrypting large amounts of data \cite{tracy2002guidelines}. To address this, key encapsulation mechanisms (KEMs) are used to securely transmit session keys for symmetric-key cryptography \cite{katz2007introduction}. The process involves a transmitter (Alice) using the recipient’s (Bob) public key to encrypt a randomly generated session key, producing a ciphertext. The recipient decrypts this ciphertext with their private key to retrieve the session key. This session key, now securely shared, is used within a symmetric cryptosystem, which is far more efficient for encrypting and decrypting large volumes of data. This method leverages the security of public-key cryptography for key exchange while benefiting from the speed and efficiency of symmetric encryption.

Current widely deployed KEMs are vulnerable to quantum attacks \cite{365700,Shor}, making them unsafe in a future where quantum computers become operational. Consequently, there is an urgent need for quantum-resistant KEMs to safeguard cryptographic systems against these emerging threats. However, many quantum-resistant algorithms are relatively new and have not undergone the extensive scrutiny of traditional (non-quantum secure) cryptographic algorithms \cite{dubrova2023breaking}. Table~\ref{tab:kem_schemes} summarizes the current status of selected KEMs that were proposed during the NIST Post-Quantum Cryptography (PQC) standardization process as of April 2025. In particular, CRYSTALS-Kyber has been officially selected and standardized \cite{kyber2022}. Classic McEliece, BIKE, and HQC are alternate finalists and remain under evaluation in Round 4 \cite{mceliece2022,bike2022,hqc2022}. SIKE as well was an alternate candidate proposed in Round 4 \cite{SIKE2022sidh}, but has since been broken via an efficient key-recovery attack \cite{castryck2022sidh}.

To address this uncertainty, there has been a demand for hybrid key exchange algorithms \cite{irtf-cfrg-hybrid-kems-03}. These algorithms combine quantum-resistant KEMs with traditional, well-established KEMs. This approach provides a safety net, allowing users to benefit from the potential security of post-quantum algorithms while retaining the longer lived reliability of traditional cryptosystems. Hybrid systems are particularly valuable in contexts where regulatory requirements, such as FIPS compliance, mandate the continued use of traditional algorithms \cite{fips203}. For users concerned about the future threat of retroactive decryption, also known as \say{Harvest now, decrypt later}, where adversaries can currently store encrypted data now which they might potentially decrypt later after a cryptographic breakthrough, hybrid key exchange offers a practical interim solution.

\begin{figure}[!t]
  \centering
    \includegraphics[width=.98 \columnwidth]{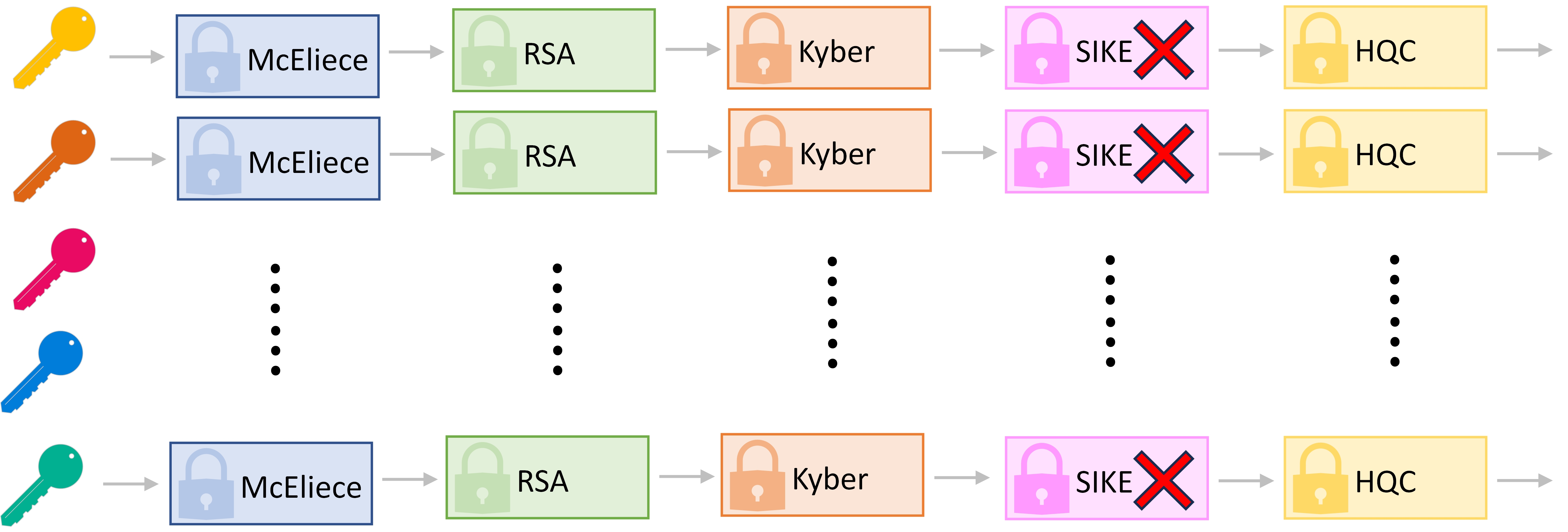}
    \includegraphics[width=.45405 \columnwidth]{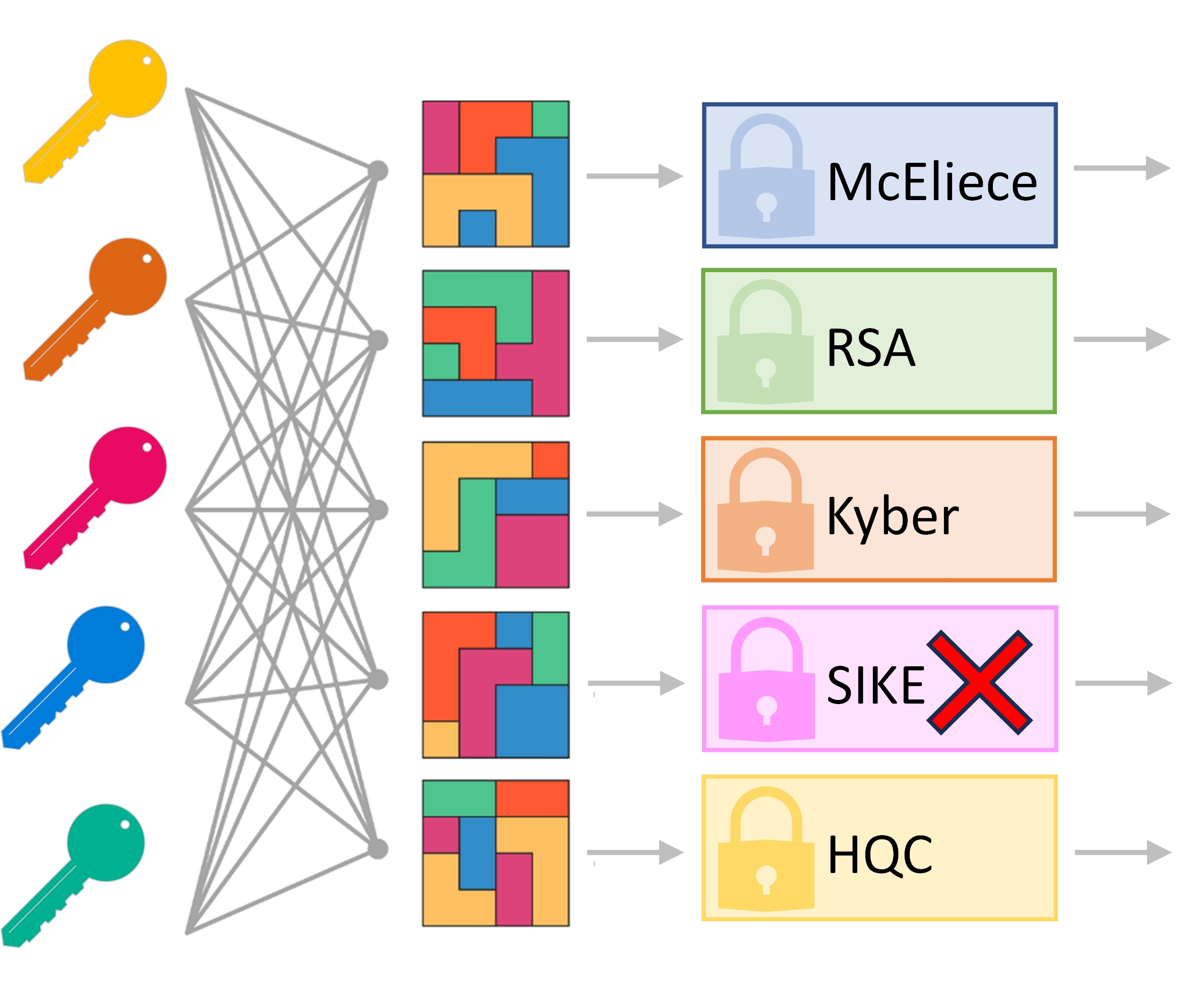}
    \vspace{-0.4cm}
  \caption{KEM schemes: Serial Encapsulation (top, see Example~\ref{eg: serial}) compared to CHOKE (bottom, see Example~\ref{eg: choke}).}
  \label{fig:KEM_Series}
\end{figure}
\begin{table}
\centering
\begin{tabular}{@{}llll@{}}
\toprule
\textbf{Algorithm} & \textbf{Status} & \textbf{Cryptographic Type} & \hspace{-0.3cm}\textbf{Ref.} \\ \midrule
CRYSTALS-Kyber     & Standardized & Lattice-based (MLWE) & \hspace{-0.3cm} \cite{kyber2022} \\
Classic McEliece   & Alternate finalist & Code-based (Goppa) & \hspace{-0.3cm} \cite{mceliece2022} \\
BIKE               & Alternate finalist & Code-based (QC-MDPC) & \hspace{-0.3cm} \cite{bike2022} \\
HQC                & Alternate finalist & Code-based (LPN-style) & \hspace{-0.3cm} \cite{hqc2022} \\
SIKE               & Broken (2022) \cite{castryck2022sidh} & Isogeny-based (SIDH) & \hspace{-0.4cm} \cite{SIKE2022sidh} \\
\bottomrule
\vspace{-0.2cm}
\end{tabular}
\caption{\verysmall Selected KEMs from the NIST PQC project (as of April 2025).}
\label{tab:kem_schemes}
\vspace{-0.8cm}
\end{table}

The primary goal of a hybrid key exchange mechanism is to establish a session key that remains secure as long as at least one of the component key exchange methods remains unbroken. A straightforward approach to achieving security in a hybrid key exchange mechanism is to concatenate KEMs together. In this scheme, the security of the session key is ensured as long as at least one KEM remains unbroken. However, this method multiplies the computational cost by the number of KEMs utilized, as it requires encrypting the data once with each KEM.

\begin{figure*}[!t]
  \centering
  \vspace{0.05cm}
  \includegraphics[width=\linewidth]{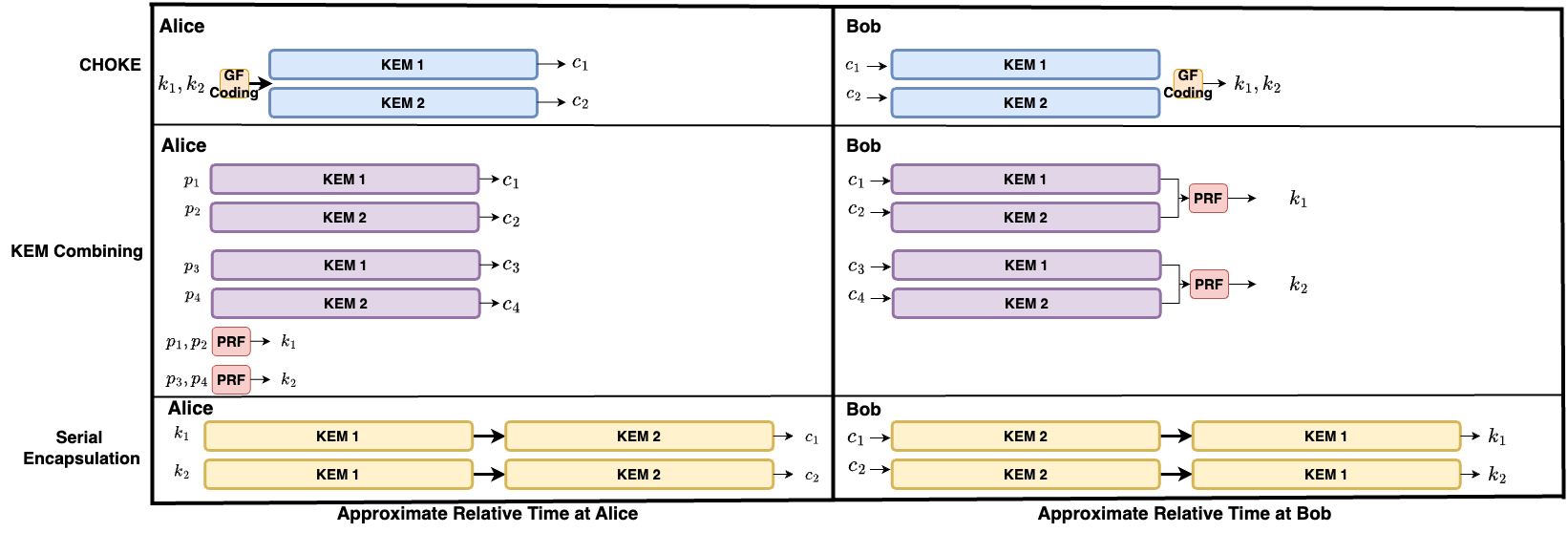}
  \vspace{-0.5cm}
  \caption{Visual comparison of computational operations for transporting two session keys using the three hybrid KEM schemes described in the Introduction. The left side represents encapsulation operations at Alice (sender), and the right side represents decapsulation operations at Bob (receiver). Block sizes illustrate relative computational costs (not to scale). Even in this basic scenario, CHOKE reduces encapsulation and decapsulation operations by half compared to the other two schemes. Generalizing to $n$ session keys, CHOKE requires only $n$ encapsulations and decapsulations, whereas the other methods require $n^2$.}
  \label{fig:eff}
  \vspace{-0.4cm}
\end{figure*}

In this paper, we present CHOKE (Code-based Hybrid Optimal Key Exchange), an efficient hybrid key encapsulation mechanism designed to transmit multiple session keys simultaneously. The protocol operates by encoding each session key into a linear combination using an individually secure code and encapsulating each encoded part with its corresponding KEM. This guarantees individual secrecy, as the adversary learns nothing about any single session key provided at least one KEM remains secure.

Our main contributions are as follows:

\begin{itemize}[wide, labelwidth=0.3cm, labelindent=1pt]
    \item In Theorem~\ref{thm:choke-cost}, we analyze the computational cost of CHOKE and show an $n$-fold reduction compared to conventional hybrid schemes. Specifically, CHOKE requires performing only one encapsulation and decapsulation per KEM, regardless of the number of session keys.

    \item In Theorem~\ref{thm:choke-comm}, we determine the exact communication cost of CHOKE, showing that the total bandwidth is the sum of ciphertext lengths from each KEM. Furthermore, in Theorem~\ref{thm:comm-optimal}, we prove this cost is optimal under the assumption that session keys are incompressible and that each KEM must be invoked at least once to securely encapsulate all the keys.

    \item In Theorem~\ref{thm:choke-security}, we prove the security of CHOKE using a simulation-based approach, showing that an adversary who breaks all but one of the underlying KEMs learns no information about any individual session key.

    \item In Section~\ref{sec: relatedkey}, we discuss potential security risks associated with related-key attacks. While CHOKE guarantees that individual session keys remain secure if at least one KEM is unbroken, breaking certain KEMs allows an adversary to learn linear combinations of keys. Therefore, we highlight the necessity of employing symmetric-key encryption schemes robust against related-key attacks, such as AES, when using keys transported by CHOKE.
\end{itemize}

\subsection{An Example with Two KEMs}

To show how our protocol works we compare it with two other schemes, simple concatenation, and the KEM Combiners scheme of \cite{giacon2018kem}.

Consider Alice wants to send two session keys, $k_1,k_2 \in \mathbb{F}_q^{d}$ chosen uniformly at random, to Bob. Because of regulations, Bob must utilize a traditionally approved public-key encryption scheme $\mathsf{KEM}_{1}$.\footnote{\label{foot:finite}To simplify our presentation, we assume throughout that we are working over a finite field $\mathbb{F}_q$ of sufficiently large size.} But because of his concerns about the future of quantum computing he also wants to utilize a post-quantum scheme $\mathsf{KEM}_{2}$. For each $\mathsf{KEM}_{i}$ we denote the encapsulation function by $\mathsf{KEM.enc}_{i}: \mathbb{F}_q^{d_i} \times \mathcal{P}_i \rightarrow \mathbb{F}_q^{\ell_i}$, the decapsulation function by $\mathsf{KEM.dec}_{i}:\mathbb{F}_q^{\ell_i} \times \mathcal{S}_i \rightarrow \mathbb{F}_q^{d_i}$, the public key by $pk_i \in \mathcal{P}_i$ and the private key by $sk_i \in \mathcal{S}_i$. We assume anyone can know the public keys, but only Bob knows the private keys.

\begin{example}{(Serial Encapsulation).} \label{eg: serial}
Alice encapsulates the session key $k_i$ with $\mathsf{KEM}_{2}$ and then encapsulates that with $\mathsf{KEM}_{1}$ to obtain the ciphertexts
\vspace{-0.2cm}
\begin{align*}
    c_1 &= \mathsf{KEM.enc}_{2} (\mathsf{KEM.enc}_{1} (k_1, pk_1),pk_2), \\
    c_2 &= \mathsf{KEM.enc}_{2} (\mathsf{KEM.enc}_{1} (k_2, pk_1),pk_2),
\end{align*}
which are then sent to Bob.

Bob utilizes the private key $sk_i$ to decapsulate the two session keys,
\vspace{-0.2cm}
\begin{align*}
k_1 &= \mathsf{KEM.dec}_{1}(\mathsf{KEM.dec}_{2}(c_1, sk_2), sk_1)\\
k_2 &= \mathsf{KEM.dec}_{1}(\mathsf{KEM.dec}_{2}(c_2, sk_2), sk_1).
\end{align*}
\begin{itemize}[wide, labelwidth=0.3cm, labelindent=1pt]
    \item \underline{Security:} In order to obtain either key, $k_1$ or $k_2$, the adversary must break both $\mathsf{KEM}_{1}$ and $\mathsf{KEM}_{2}$.

    \item \underline{Computation cost:} Each key must be encapsulated twice. Thus, the computational cost is that of performing two $\mathsf{KEM}_{1}$ operations plus two $\mathsf{KEM}_{2}$ operations.

    \item \underline{Communication cost:} Alice transmits the two outer ciphertexts $(c_1,c_2)$ produced by $\mathsf{KEM}_1$. Each ciphertext lies in $\mathbb{F}_q^{\ell_1}$, so the total bandwidth is $2\ell_1$ symbols of $\mathbb{F}_q$.
\end{itemize}

\end{example}

\begin{example}{(KEM Combining).} \label{eg: KEM_combiner}
Alice generates four random symbols $p_1,p_2,p_3,p_4$. The session keys are created utilizing a pseudorandom function, $k_1 = \mathsf{PRF} (p_1,p_2)$ and $k_2 = \mathsf{PRF} (p_3,p_4)$. Alice performs the following encapsulations
\vspace{-0.1cm}
\small
\begin{align*}
c_1 &= \mathsf{KEM.enc}_{1}(p_1,pk_1), \quad c_2 = \mathsf{KEM.enc}_{2}\small{(p_2,pk_2)}, \\
c_3 &= \mathsf{KEM.enc}_{1}(p_3,pk_1), \quad c_4 = \mathsf{KEM.enc}_{2}(p_4,pk_2),
\end{align*}
\normalsize
and sends $c_1,c_2,c_3,c_4$ to Bob.
 
Bob decapsulates the four ciphertexts
\vspace{-0.2cm}
\begin{align*}
p_1 &= \mathsf{KEM.dec}_{1}(c_1,sk_1), \quad p_2 = \mathsf{KEM.dec}_{2}(c_2,sk_2),   \\     
p_3 &= \mathsf{KEM.dec}_{1}(c_3,sk_1), \quad p_4 = \mathsf{KEM.dec}_{2}(c_4,sk_2),
\end{align*}
to obtain $p_1,p_2,p_3,p_4$ and reconstructs the session keys $k_1 = \mathsf{PRF}(p_1,p_2)$ and $k_2 = \mathsf{PRF}(p_3,p_4)$.

\begin{itemize}[wide, labelwidth=0.3cm, labelindent=1pt]
\item \underline{Security:} To recover a session key, say $k_1$, an adversary must obtain both $p_1$ and $p_2$.  Each
block is protected by a different KEM, so breaking a single
$\mathsf{KEM}_{i}$ is insufficient.

\item \underline{Computation Cost:} The scheme performs two $\mathsf{KEM}_{1}$ and two $\mathsf{KEM}_{2}$ operations -- the same computational complexity as in Example~\ref{eg: serial}. However, all four can execute in parallel. Hence, the time complexity is reduced to one encapsulation (or decapsulation) of the slower KEM, rather than the sum of all four.

\item \underline{Communication cost:}  
Alice sends four ciphertexts, $(c_1,c_3)$ produced by $\mathsf{KEM}_1$ and $(c_2,c_4)$ produced by $\mathsf{KEM}_2$. Hence the total bandwidth is $2\ell_{1}+2\ell_{2}$ symbols of $\mathbb{F}_{q}$, whereas the serial construction of Example~\ref{eg: serial} needs only $2\ell_{1}$ symbols.

\end{itemize}

\end{example}

\begin{example}{(CHOKE).} \label{eg: choke}
Alice concatenates the two session keys $K=[k_1,k_2]$ into a vector and multiplies it with a public generator matrix $\mathbf{G}=%
\bigl(\begin{smallmatrix}1&1\\ 2&1\end{smallmatrix}\bigr)\!\in\!\mathbb{F}_{q^{u}}^{2\times2}$,
obtaining
\[
X=K\mathbf{G}=[k_1+k_2,k_1+2k_2]=[X_1,X_2].
\]
Then, she encapsulates each linear combination and sends the ciphertexts to Bob:
\vspace{-0.2cm}
\[
c_1=\mathsf{KEM.enc}_{1}(X_1,pk_1),\qquad
c_2=\mathsf{KEM.enc}_{2}(X_2,pk_2).
\]

Bob decapsulates
\vspace{-0.2cm}
\[
X_1=\mathsf{KEM.dec}_{1}(c_1,sk_1),\qquad
X_2=\mathsf{KEM.dec}_{2}(c_2,sk_2),
\]
and recovers the original session keys $[k_1,k_2]=\mathbf{G}^{-1}X$.

\begin{itemize}[wide, labelwidth=0.3cm, labelindent=1pt]
  \item \underline{Security:} Both ciphertexts are required to reconstruct $K$. If an adversary breaks one KEM he learns a linear combination of the session keys. In Theorem~\ref{thm:choke-security} we show that this reveals no information about any individual session key, i.e., each session key $k_i$ is computationally indistinguishable from a uniform random one.
  
  \item \underline{Computation Cost:} The scheme performs a single $\mathsf{KEM}_{1}$ and $\mathsf{KEM}_{2}$ operation -- halving the computation cost when compared with the schemes in Examples~\ref{eg: serial} and \ref{eg: KEM_combiner}.

  \item \underline{Communication Cost:}  
Only two ciphertexts are sent—$c_{1}$ of length $\ell_{1}$ and $c_{2}$ of length $\ell_{2}$.  The total bandwidth is therefore $\ell_{1}+\ell_{2}$ symbols of $\mathbb{F}_{q}$. This is less than both the symbols required by the serial scheme of Example~\ref{eg: serial} and by KEM Combining in Example~\ref{eg: KEM_combiner}.
\end{itemize}
\end{example}

\subsection{Related Work}

Secret sharing was originally proposed by Blakley and Shamir as a technique for safeguarding cryptographic keys. Karnin et al. \cite{hellman2} extended this concept by showing that when sharing multiple cryptographic keys simultaneously, performance can be significantly improved through the notion of individual security \cite{Hellman}, where each key remains independently secure. Individual security has since been successfully applied across diverse communication and storage domains. Notable examples include single communication links \cite{SCMUniform}, broadcast channels \cite{mansour2014secrecy,chen2015individual,mansour2015individual,mansour2016individual}, multiple-access channels \cite{goldenbaum2015multiple,chen2016secure}, networks and multicast communications \cite{bhattad2005weakly,silva2009universal,cohen2018secure}, algebraic security schemes \cite{lima2007random,claridge2017probability}, terahertz wireless systems \cite{cohen2023absolute}, angularly dispersive optical links \cite{yeh2023securing}, and distributed storage systems \cite{kadhe2014weakly,kadhe2014weakly1,paunkoska2016improved,paunkoska2018improving,bian2019optimal}. Individual security ensures that an eavesdropper obtaining any limited subset of the shared information learns no useful information about each message individually, although they may acquire some insignificant, controlled leakage about combinations of the messages.

Our work builds upon the framework proposed by Cohen et al. \cite{cohen2021network}, where Hybrid Universal Network-Coded Cryptography (HUNCC) was introduced to enhance the efficiency of cryptographic systems through partial encryption. Our scheme can be viewed as a specialized instance of HUNCC tailored explicitly for transmitting cryptographic keys, where we instead utilize full encryption for each coded symbol. If certain underlying KEMs become compromised, the security assurances of our scheme directly parallel those established by HUNCC under analogous partial encryption conditions. In this paper, we also introduce a new proof of security using the real-world/ideal-world (simulationist) paradigm, which can readily extend to provide an alternative, simulation-based security proof for the general HUNCC framework \cite{cohen2022partial}.


\section{Security Definitions}\label{sec:prelim}

In this section we introduce security definitions we utilize in the paper. All honest parties and adversaries are modeled as probabilistic algorithms whose running time is bounded by a polynomial in the global security parameter~$\kappa$.

\begin{definition}[Negligible function]
A function $f:\mathbb{N}\to\mathbb{R}_{\ge0}$ is negligible if for every constant $c>0$ there exists $N$ such that for all $\kappa>N$, $f(\kappa)<\kappa^{-c}$.
\end{definition}

\begin{definition}[Probabilistic polynomial‐time (PPT)]
An algorithm is probabilistic polynomial‐time (PPT) if its running time is bounded by a polynomial in the length of its input.
\end{definition}

A key encapsulation mechanism (KEM) enables two parties to securely establish a session key over an insecure channel.

\begin{definition}[Key Encapsulation Mechanism over $\mathbb{F}_{q}$]\hspace{-0.25cm}$^{\ref{foot:finite}}$\label{def:kem-fq}
Fix a security parameter $\kappa\in\mathbb{N}$ and let $q=q(\kappa)$ be a prime power bounded by $\operatorname{poly}(\kappa)$. For integers $d=d(\kappa)$ and $\ell=\ell(\kappa)$ define the key space $\mathcal{K}=\mathbb{F}_{q}^{d}$ and the ciphertext space $\mathcal{C}=\mathbb{F}_{q}^{\ell}$. A key–encapsulation mechanism (KEM) is a triple of PPT algorithms $\mathsf{KEM}=(\mathsf{KEM.gen},\mathsf{KEM.enc},\mathsf{KEM.dec})$
with public/secret key sets $\mathcal{P}$ and $\mathcal{S}$ that satisfy:
\begin{enumerate}
  \item \underline{Key generation:} $(\mathsf{pk},\mathsf{sk})\leftarrow\mathsf{Gen}(1^{\kappa})$.
  \item \underline{Encapsulation:} given $\mathsf{pk}\in\mathcal{P}$ and a uniformly random session key $m \in \mathcal{K}$, output $c:=\mathsf{KEM.enc}(m,\mathsf{pk})\in\mathcal{C}$.
  \item \underline{Decapsulation:} given $\mathsf{sk}\in\mathcal{S}$ and $c\in\mathcal{C}$, output $m':=\mathsf{KEM.dec}(c,\mathsf{sk})\in\mathcal{K}$.
\end{enumerate}
\underline{Correctness:}
There exists a negligible function $\mathsf{negl}(\cdot)$ such that for every $\kappa$,
\begin{align*}
  \Pr \bigl[ \mathsf{KEM.dec}\bigl(\mathsf{KEM.enc}(m,\mathsf{pk}), \mathsf{sk}\bigr)=m]\ge 1-\mathsf{negl}(\kappa).
\end{align*}
\end{definition}

We now formalize the standard IND‑CPA security notion a real‑vs‑ideal experiment.

\begin{definition}[IND‑CPA security via real/ideal experiments]\label{def:ind-cpa-fq}
Fix a KEM  with key space $\mathcal{K}=\mathbb{F}_{q}^{\,d(\kappa)}$ and ciphertext space $\mathcal{C}=\mathbb{F}_{q}^{\,\ell(\kappa)}$. We say that $\mathsf{KEM}$ is indistinguishable under a chosen‑plaintext
attack (IND‑CPA) if for every PPT distinguisher $\mathcal{D}$ there exists a negligible function $\mathsf{negl}(\kappa)$ such that, for all security parameters $\kappa$ and for all $m,m'\in\mathcal{K}$, the two experiments below differ in $\mathcal{D}$’s acceptance probability by at most $\mathsf{negl}(\kappa)$.

\medskip
\fbox{\parbox{.96\columnwidth}{
\textbf{Real experiment} $\mathsf{Real}_{m}(\kappa)$
\begin{enumerate}
  \item $(\mathsf{pk},\mathsf{sk})\leftarrow\mathsf{KEM.gen}(1^{\kappa})$;
  \item $c\leftarrow\mathsf{KEM.enc}(m,\mathsf{pk})$;
  \item output $\,\mathcal{D}(c,\mathsf{pk})$.
\end{enumerate}
}}

\medskip
\fbox{\parbox{.96\columnwidth}{
\textbf{Ideal experiment} $\mathsf{Ideal}_{m'}(\kappa)$
\begin{enumerate}
  \item $(\mathsf{pk},\mathsf{sk})\leftarrow\mathsf{KEM.gen}(1^{\kappa})$;
  \item $c^*\leftarrow\mathsf{KEM.enc}(m',\mathsf{pk})$;
  \item output $\,\mathcal{D}(c^*,\mathsf{pk})$.
\end{enumerate}
}}

\medskip
Formally,
\[
  \bigl|\Pr[\mathcal{D}(\mathsf{Real}_{m}(\kappa))=1]
        -
        \Pr[\mathcal{D}(\mathsf{Ideal}_{m'}(\kappa))=1]\bigr|
  \le \mathsf{negl}(\kappa).
\]
\end{definition}

In Theorem~\ref{thm:choke-security} we show that if all KEMs in CHOKE's algorithm besides one are compromised, no information is leaked about any individual key. The proof works by showing that a computationally bounded adversary cannot distinguish between that setting and an information-theoretically secure setting with individual security.

\vspace{-0.18cm}
\begin{algorithm} \label{alg:choke}
\caption{CHOKE: Code‑based Hybrid KEM}\label{alg:choke}
\begin{algorithmic}[1]
\Statex \textbf{Key generation at Bob}
\For{$i \gets 1$ \textbf{to} $n$}
  \State $(pk_i,sk_i) \gets \mathsf{KEM.gen}_i$
\EndFor
\State publish $(pk_1,\ldots,pk_n)$
\vspace{4pt}

\Statex \textbf{Encapsulation at Alice}
\State \textbf{Input:} session keys $K=[k_1,\ldots,k_n]\in\mathbb{F}_q^{n}$
\State $X \gets K \cdot G$ \Comment{$G$ is the public generator matrix}
\For{$i \gets 1$ \textbf{to} $n$}
  \State $c_i \gets \mathsf{KEM.enc}_i(X_i,pk_i)$
\EndFor
\State send $(c_1,\ldots,c_n)$ to Bob
\vspace{4pt}

\Statex \textbf{Decapsulation at Bob}
\For{$i \gets 1$ \textbf{to} $n$}
  \State $X_i \gets \mathsf{KEM.dec}_i(c_i,sk_i)$
\EndFor
\State $K \gets G^{-1} \cdot X$ \Comment{recover $[k_1,\ldots,k_n]$}
\State \Return $K$
\end{algorithmic}
\end{algorithm}

\begin{definition}[Individual security]
An encoding of $n$ independent uniform messages $M_1,\dots,M_n$ into outputs $X_1,\dots,X_n$ is individually secure if for every subset of at most $n-1$ outputs, say $(X_{i_1},\dots,X_{i_{n-1}})$, and for each index $j\in\{1,\dots,n\}$,
\[
  I\bigl(M_j;\,X_{i_1},\dots,X_{i_{n-1}}\bigr)=0,
\]
where $I(\cdot;\cdot)$ denotes the mutual information.
\end{definition}

\section{CHOKE Algorithm}\label{proposed_scheme}
\vspace{-0.1cm}
We now present our construction, CHOKE, as Algorithm~\ref{alg:choke}. In our setting, Alice wants to send $n$ session keys $k_1,\ldots,k_n$ to Bob. Bob has $n$ key encapsulation mechanisms $\mathsf{KEM}_1, \ldots, \mathsf{KEM}_n$ with corresponding public and private keys $(pk_1,sk_1),\ldots,(pk_n,sk_n)$. The public keys are made public.

\underline{Encapsulation:} Alice concatenates the session keys into a vector $K=[k_1,\ldots,k_n]$ and multiplies it with a public generator matrix $G$ of an individually secure code to obtain $KG = [X_1,\ldots,X_n]$. Then, she encapsulates each linear combination into the ciphertext $c_i = \mathsf{KEM.enc}_i (X_i,pk_i)$ and sends it to Bob.

\underline{Decapsulation:} Bob decapsulates each ciphertext $c_i$ to obtain $X_i = \mathsf{KEM.dec}_i (\mathsf{KEM.enc}_i (X_i,pk_i),sk_i)$ and then recovers the session keys $[k_1,\ldots,k_n]=G^{-1} [X_1,\ldots,X_n]$.

\section{Performance}
\vspace{-0.1cm}
In this section, we analyze both the computation and communication cost of CHOKE. Moreover, we show that CHOKE achieves an optimal communication cost.

\subsection{Computation Cost}

\begin{theorem}[Computation Cost]\label{thm:choke-cost}
Let $\mathsf{KEM}_1,\dots,\mathsf{KEM}_n$ be the mechanisms used in CHOKE, and denote by $E_i$ and $D_i$ the computational cost of calling $\mathsf{KEM.enc}_i$ and
$\mathsf{KEM.dec}_i$, respectively. Then, the computational cost at Alice and Bob is $\sum_{i=1}^{n} E_i$ and $\sum_{i=1}^{n} D_i$, respectively.
\end{theorem}

\begin{proof}
Algorithm~\ref{alg:choke} encapsulates each coded block $X_i$ once and decapsulates the corresponding ciphertext $c_i$ once.  Hence $\mathsf{KEM}_i$ is invoked exactly one time for encapsulation and one time for decapsulation.
\end{proof}

In both alternative hybrids every one of the $n$ session keys is handled by all underlying mechanisms: once when Alice encapsulates the key and once when Bob decapsulates it.  
In the Serial Encapsulation scheme (exemplified in Example~\ref{eg: serial}) every mechanism $\mathsf{KEM}_i$ is executed $n$ times across the $n$ keys, giving Alice a total cost of\vspace{-0.15cm}
\[\vspace{-0.15cm}
   \sum_{j=1}^{n}\sum_{i=1}^{n} E_i \;=\; n\!\sum_{i=1}^{n} E_i,
   \qquad
   \text{and Bob a cost of } 
   n\!\sum_{i=1}^{n} D_i.
\]
The KEM‑Combining construction (exemplified in Example~\ref{eg: KEM_combiner}) performs its encapsulations in parallel rather than in series, yet it still invokes each $\mathsf{KEM}_i$ once per key; hence its transmit‑side and receive‑side costs are the same as in the serial scheme, namely $n\sum_{i=1}^{n}E_i$ and $n\sum_{i=1}^{n}D_i$.



Since \textsc{CHOKE} protects all $n$ keys with a single invocation of each mechanism, only $\sum_{i=1}^{n} E_i$ and $\sum_{i=1}^{n} D_i$ encapsulation and decapsulation operations are performed by Alice and Bob respectively. Thus \textsc{CHOKE} achieves an $n$‑fold reduction in computational effort at both ends of the channel compared with either conventional hybrid approach.

\subsection{Communication Cost}

\begin{theorem}[Communication cost]\label{thm:choke-comm}
Let $\mathsf{KEM}_1,\dots,\mathsf{KEM}_n$ be the mechanisms used in \textsc{CHOKE}, and let $\mathbb{F_q}^\ell_i$ be the output space of each $\mathsf{KEM.enc}_i$.  Then, the communication cost of \textsc{CHOKE} is $\sum_{i=1}^{n}\ell_i$ bits.
\end{theorem}

\begin{proof}
Algorithm~\ref{alg:choke} outputs the tuple $(c_1,\dots,c_n)$ with each $|c_i|=\ell_i$ and transmits it once.
\end{proof}

In the Serial Encapsulation scheme (exemplified in Example~\ref{eg: serial}) every session key is encapsulated by the outermost mechanism $\mathsf{KEM}_1$; consequently Alice emits $n$ ciphertexts, each of length at most $\ell_{\max}:=\max_i\ell_i$, so the communication volume is $n\ell_{\max}$ bits. The KEM‑Combining scheme (exemplified in Example~\ref{eg: KEM_combiner}) is more demanding. Because each key is protected by all mechanisms, the sender must transmit $n$ ciphertexts for every $\mathsf{KEM}_i$, giving a total of $n\sum_{i=1}^{n}\ell_i$ bits.

\textsc{CHOKE}, by contrast, needs only the $\sum_{i}\ell_i$ bits stated in Theorem~\ref{thm:choke-comm}. Thus its bandwidth is strictly smaller than that of KEM-Combining by a factor of $\approx n$, and it never exceeds the serial scheme’s cost.

We now show that CHOKE achieves optimality in terms of communication cost. Specifically, we demonstrate that any hybrid KEM scheme that securely encapsulates $n$ independent session keys and invokes each $\mathsf{KEM}_i$ at least once must transmit at least as many symbols as CHOKE does, namely $\sum_{i=1}^n \ell_i$ symbols of $\mathbb{F}_q$.

\begin{theorem}[Optimality of CHOKE]\label{thm:comm-optimal}
Let $\mathsf{KEM}_1,\dots,\mathsf{KEM}_n$ be mechanisms whose
ciphertext lengths are $\ell_1,\dots,\ell_n$. Consider a hybrid protocol that, on input a tuple of session keys $k_1,\dots,k_n\in\mathcal{K}$, must invoke each $\mathsf{KEM}_i$ at least once. Then, the protocol must transmit at least $\sum_{i=1}^n\ell_i$ symbols of $\mathbb{F}_{q}$.
\end{theorem}

\begin{proof}
By correctness, Bob must obtain each ciphertext $c_i\in\mathbb{F}_{q}^{\ell_i}$ exactly as produced by
$\mathsf{KEM}_i.\mathrm{enc}$, since $\mathsf{KEM}_i.\mathrm{dec}$ may fail on any other string.
Hence the message $M$ that Alice sends in the protocol must allow Bob to recover the tuple $(c_1,\dots,c_n)$ without error.   Formally, the map $\mathcal{E}:\,(c_1,\dots,c_n)\rightarrow M$ implemented by the protocol must be injective. Otherwise, two distinct ciphertext tuples would be mapped to the same $M$, and Bob could not determine which tuple to decapsulate, contradicting correctness.

Because $\mathcal{E}$ is injective, $|M|$ (the number of $\mathbb{F}_{q}$ symbols in~$M$) must be at least the length of the concatenation $c_1\,\|\,\dots\,\|\,c_n$, which equals $\sum_{i=1}^{n}\ell_i$. Therefore every correct protocol that invokes each KEM once necessarily transmits at least $\sum_{i=1}^{n}\ell_i$ symbols.
\end{proof}
\vspace{-0.4cm}
\section{Security}
We now show that CHOKE retains the desired hybrid KEM property, namely that if all but one of the underlying KEMs are compromised, the adversary learns no information about any individual session key. Our proof uses a standard simulation-based argument \cite{lindell2017simulate}: we demonstrate that no computationally bounded adversary can distinguish between CHOKE's ciphertexts and an idealized scenario in which the ciphertexts achieve information-theoretic individual security.

\begin{theorem}[Individual key IND security]\label{thm:choke-security}
Suppose $\mathsf{KEM}_i$ is IND‐CPA secure.  Then in the \textsc{CHOKE} protocol, even if an adversary Eve breaks all KEMs except for $\mathsf{KEM}_i$, no PPT distinguisher can learn any information about any individual session key $k_i$ except with negligible probability in the security parameter~$\kappa$.
\end{theorem}

\begin{proof}
We prove the claim by a standard real‐vs‐ideal simulation argument.  Fix
any PPT adversary Eve.  Let $(k_1,\dots,k_n)$ be the uniformly random
session keys and let
\[
X = G\,[k_1;\dots;k_n] = (X_1,X_2,\dots,X_n),
\]
be the coded linear combinations.  

Without loss of generality, suppose that Eve breaks all the KEMs but the first one. Then, in the real execution, Eve’s view is $\mathsf{Real} =\bigl(c_1,X_2,\dots,X_n\bigr)$, where $c_1 = \mathsf{KEM.enc}_1(X_1,pk_1)$. Define an ideal execution in which Eve instead sees $\mathsf{Ideal} =\bigl(c_1^*,X_2,\dots,X_n\bigr)$ where $c_1^* =  \mathsf{KEM.enc}_1(0,pk_1)$. Because $\mathsf{KEM}_1$ is IND‐CPA secure, for every PPT distinguisher $\mathcal{D}$ there is a negligible function $\mathsf{negl}(\kappa)$ such that
\[
  \bigl|\Pr[\mathcal{D}(\mathsf{Real})=1]
       -
       \Pr[\mathcal{D}(\mathsf{Ideal})=1]\bigr|
  \le
  \mathsf{negl}(\kappa).
\]
In particular this holds for any adversary Eve attempting to  distinguish the two executions.

We observe that in the ideal execution $c_1^*$ is independent of $X_1,\dots,X_n$.  Since $G$ is individually secure, it follows that $I\bigl(k_i ;\mathsf{Ideal}\bigr) = 0$ for every $i=1,\dots,n$. Thus, Eve gains no information about any individual key in the ideal world. But by the IND‐CPA bound above, Eve’s view in the real world is computationally indistinguishable from her view in the ideal world. Therefore, Eve’s advantage in learning any information about any individual $k_i$ in the real view is at most $\mathsf{negl}(\kappa)$.
\end{proof}

\section{CHOKE and Related-Key Attacks} \label{sec: relatedkey}

In CHOKE, if an underlying KEM is compromised, even though no information about any individual key is leaked, the adversary learns linear combinations of the session keys. When these session keys are subsequently used in symmetric-key encryption schemes, the exposure of linear relationships between the keys can potentially make the encryption vulnerable to related-key attacks~\cite{biham1994new}. 

Although modern symmetric-key encryption schemes, such as AES, are explicitly designed to be robust against related-key attacks~\cite{biryukov2009related,biryukov2009distinguisher,bellare2003theoretical}, there have been theoretical scenarios where knowledge of key relationships increased the adversary's success probability~\cite{biham2005related,roetteler2015note}.

This vulnerability is not unique to CHOKE; it also arises in other multi-secret sharing protocols. Multi-secret sharing schemes inherently leak correlations between secrets, similarly exposing the system to related-key attacks. Therefore, when deploying CHOKE, it is crucial to ensure that the symmetric encryption scheme utilizing the transported keys is resilient to related-key attacks.

\bibliographystyle{IEEEtran}
\bibliography{references.bib}

\begin{thebibliography}{10}
\providecommand{\url}[1]{#1}
\csname url@samestyle\endcsname
\providecommand{\newblock}{\relax}
\providecommand{\bibinfo}[2]{#2}
\providecommand{\BIBentrySTDinterwordspacing}{\spaceskip=0pt\relax}
\providecommand{\BIBentryALTinterwordstretchfactor}{4}
\providecommand{\BIBentryALTinterwordspacing}{\spaceskip=\fontdimen2\font plus
\BIBentryALTinterwordstretchfactor\fontdimen3\font minus \fontdimen4\font\relax}
\providecommand{\BIBforeignlanguage}[2]{{%
\expandafter\ifx\csname l@#1\endcsname\relax
\typeout{** WARNING: IEEEtran.bst: No hyphenation pattern has been}%
\typeout{** loaded for the language `#1'. Using the pattern for}%
\typeout{** the default language instead.}%
\else
\language=\csname l@#1\endcsname
\fi
#2}}
\providecommand{\BIBdecl}{\relax}
\BIBdecl

\bibitem{tracy2002guidelines}
M.~Tracy, W.~Jansen, and M.~McLarnon, ``Guidelines on securing public web servers,'' \emph{NIST Special Publication}, vol. 800, p.~44, 2002.

\bibitem{katz2007introduction}
J.~Katz and Y.~Lindell, \emph{Introduction to modern cryptography: principles and protocols}.\hskip 1em plus 0.5em minus 0.4em\relax Chapman and hall/CRC, 2007.

\bibitem{365700}
P.~Shor, ``Algorithms for quantum computation: discrete logarithms and factoring,'' in \emph{Proceedings 35th Annual Symposium on Foundations of Computer Science}, 1994, pp. 124--134.

\bibitem{Shor}
\BIBentryALTinterwordspacing
P.~W. Shor, ``Polynomial-time algorithms for prime factorization and discrete logarithms on a quantum computer,'' \emph{SIAM Journal on Computing}, vol.~26, no.~5, pp. 1484--1509, 1997. [Online]. Available: \url{https://doi.org/10.1137/S0097539795293172}
\BIBentrySTDinterwordspacing

\bibitem{dubrova2023breaking}
E.~Dubrova, K.~Ngo, J.~G{\"a}rtner, and R.~Wang, ``Breaking a fifth-order masked implementation of crystals-kyber by copy-paste,'' in \emph{Proceedings of the 10th ACM Asia Public-Key Cryp. Workshop}, 2023, pp. 10--20.

\bibitem{kyber2022}
\BIBentryALTinterwordspacing
R.~Avanzi, J.~W. Bos, L.~Ducas, E.~Kiltz, T.~Lepoint, V.~Lyubashevsky, J.~M. Schanck, P.~Schwabe, G.~Seiler, and D.~Stehlé, ``{CRYSTALS-Kyber: Algorithm Specifications and Supporting Documentation},'' 2022, {NIST PQC Round 3 finalist, now standardized}. [Online]. Available: \url{https://csrc.nist.gov/Projects/post-quantum-cryptography/selected-algorithms-2022#kyber}
\BIBentrySTDinterwordspacing

\bibitem{mceliece2022}
\BIBentryALTinterwordspacing
D.~J. Bernstein, T.~Lange, and P.~Schwabe, ``{Classic McEliece: Submission and Documentation},'' 2022, {NIST PQC Round 4 alternate finalist}. [Online]. Available: \url{https://csrc.nist.gov/Projects/post-quantum-cryptography/round-4-submissions}
\BIBentrySTDinterwordspacing

\bibitem{bike2022}
\BIBentryALTinterwordspacing
P.~S. L.~M. Barreto, R.~Misoczki, J.-P. Tillich \emph{et~al.}, ``{BIKE: Bit Flipping Key Encapsulation},'' 2022, {NIST PQC Round 4 alternate finalist}. [Online]. Available: \url{https://csrc.nist.gov/Projects/post-quantum-cryptography/round-4-submissions}
\BIBentrySTDinterwordspacing

\bibitem{hqc2022}
\BIBentryALTinterwordspacing
P.~Gaborit, A.~Hauteville, J.-P. Tillich, and G.~Zémor, ``{HQC: Hamming Quasi-Cyclic Key Encapsulation Mechanism},'' 2022, {NIST PQC Round 4 alternate finalist}. [Online]. Available: \url{https://csrc.nist.gov/Projects/post-quantum-cryptography/round-4-submissions}
\BIBentrySTDinterwordspacing

\bibitem{SIKE2022sidh}
\BIBentryALTinterwordspacing
D.~J. et~al, ``{SIKE – Supersingular Isogeny Key Encapsulation},'' 2022. [Online]. Available: \url{https://sike.org/}
\BIBentrySTDinterwordspacing

\bibitem{castryck2022sidh}
\BIBentryALTinterwordspacing
W.~Castryck and T.~Decru, ``{An efficient key recovery attack on SIDH},'' \emph{Cryptology ePrint Archive}, no. 975, 2022. [Online]. Available: \url{https://eprint.iacr.org/2022/975}
\BIBentrySTDinterwordspacing

\bibitem{irtf-cfrg-hybrid-kems-03}
\BIBentryALTinterwordspacing
D.~Connolly, ``{Hybrid PQ/T Key Encapsulation Mechanisms},'' Internet Engineering Task Force, Internet-Draft draft-irtf-cfrg-hybrid-kems-03, Feb. 2025, work in Progress. [Online]. Available: \url{https://datatracker.ietf.org/doc/draft-irtf-cfrg-hybrid-kems/03/}
\BIBentrySTDinterwordspacing

\bibitem{fips203}
{\relax National Institute of Standards and Technology}, ``Module-lattice-based key-encapsulation mechanism standard.'' U.S. Department of Commerce, Washington, D.C., Tech. Rep. Federal Information Processing Standards Publications (FIPS) 203, 2024.

\bibitem{giacon2018kem}
F.~Giacon, F.~Heuer, and B.~Poettering, ``Kem combiners,'' in \emph{Public-Key Cryptography--PKC 2018: 21st IACR International Conference on Practice and Theory of Public-Key Cryptography, Rio de Janeiro, Brazil, March 25-29, 2018, Proceedings, Part I 21}.\hskip 1em plus 0.5em minus 0.4em\relax Springer, 2018, pp. 190--218.

\bibitem{hellman2}
E.~D. Karnin, J.~W. Greene, and M.~E. Hellman, ``On secret sharing systems,'' \emph{{IEEE} Trans. Inform. Theory}, vol.~29, no.~1, pp. 35--41, Jan. 1983.

\bibitem{Hellman}
A.~Carleial and M.~E. Hellman, ``A note on wyner's wiretap channel (corresp.),'' \emph{{IEEE} Trans. Inform. Theory}, vol.~23, no.~3, pp. 387--390, May 1977.

\bibitem{SCMUniform}
D.~Kobayashi, H.~Yamamoto, and T.~Ogawa, ``Secure multiplex coding attaining channel capacity in wiretap channels,'' \emph{IEEE Trans. on Inf. Theory}, vol.~59, no.~12, pp. 8131--8143, 2013.

\bibitem{mansour2014secrecy}
A.~S. Mansour, R.~F. Schaefer, and H.~Boche, ``Secrecy measures for broadcast channels with receiver side information: Joint vs individual,'' in \emph{2014 IEEE Inf. Theory Works. (ITW 2014)}.\hskip 1em plus 0.5em minus 0.4em\relax IEEE, 2014, pp. 426--430.

\bibitem{chen2015individual}
Y.~Chen, O.~O. Koyluoglu, and A.~Sezgin, ``On the individual secrecy rate region for the broadcast channel with an external eavesdropper,'' in \emph{2015 IEEE Int. Symp. on Inf. Theory (ISIT)}.\hskip 1em plus 0.5em minus 0.4em\relax IEEE, 2015, pp. 1347--1351.

\bibitem{mansour2015individual}
A.~S. Mansour, R.~F. Schaefer, and H.~Boche, ``The individual secrecy capacity of degraded multi-receiver wiretap broadcast channels,'' in \emph{2015 IEEE Int. Conf. on Comm. (ICC)}.\hskip 1em plus 0.5em minus 0.4em\relax IEEE, 2015, pp. 4181--4186.

\bibitem{mansour2016individual}
------, ``On the individual secrecy capacity regions of the general, degraded, and gaussian multi-receiver wiretap broadcast channel,'' \emph{IEEE Trans. on Inf. Fore. and Sec.}, vol.~11, no.~9, pp. 2107--2122, 2016.

\bibitem{goldenbaum2015multiple}
M.~Goldenbaum, R.~F. Schaefer, and H.~V. Poor, ``The multiple-access channel with an external eavesdropper: Trusted vs. untrusted users,'' in \emph{2015 49th Asilomar Conference on Signals, Systems and Computers}.\hskip 1em plus 0.5em minus 0.4em\relax IEEE, 2015, pp. 564--568.

\bibitem{chen2016secure}
Y.~Chen, O.~O. Koyluoglu, and A.~H. Vinck, ``On secure communication over the multiple access channel,'' in \emph{2016 Int. Symp. on Inf. Theory and Its Applications (ISITA)}.\hskip 1em plus 0.5em minus 0.4em\relax IEEE, 2016, pp. 350--354.

\bibitem{bhattad2005weakly}
K.~Bhattad and K.~R. Narayanan, ``Weakly secure network coding,'' \emph{NetCod, Apr}, vol. 104, 2005.

\bibitem{silva2009universal}
D.~Silva and F.~R. Kschischang, ``Universal weakly secure network coding,'' in \emph{2009 IEEE Inf. Theory Works. on Networking and Information Theory}.\hskip 1em plus 0.5em minus 0.4em\relax IEEE, 2009, pp. 281--285.

\bibitem{cohen2018secure}
A.~Cohen, A.~Cohen, M.~Medard, and O.~Gurewitz, ``Secure multi-source multicast,'' \emph{IEEE Trans. on Comm.}, vol.~67, no.~1, pp. 708--723, 2018.

\bibitem{lima2007random}
L.~Lima, M.~M{\'e}dard, and J.~Barros, ``Random linear network coding: A free cipher?'' in \emph{2007 IEEE Int. Symp. on Inf. Theory}.\hskip 1em plus 0.5em minus 0.4em\relax IEEE, 2007, pp. 546--550.

\bibitem{claridge2017probability}
J.~Claridge and I.~Chatzigeorgiou, ``Probability of partially decoding network-coded messages,'' \emph{IEEE Communications Letters}, vol.~21, no.~9, pp. 1945--1948, 2017.

\bibitem{cohen2023absolute}
A.~Cohen, R.~G.~L. D'Oliveira, C.-Y. Yeh, H.~Guerboukha, R.~Shrestha, Z.~Fang, E.~Knightly, M.~M{\'e}dard, and D.~M. Mittleman, ``Absolute security in terahertz wireless links,'' \emph{IEEE Journal of Selected Topics in Signal Processing}, 2023.

\bibitem{yeh2023securing}
C.-Y. Yeh, A.~Cohen, R.~G.~L. D’Oliveira, M.~M{\'e}dard, D.~M. Mittleman, and E.~W. Knightly, ``Securing angularly dispersive terahertz links with coding,'' \emph{IEEE Trans. on Inf. Forensics and Security}, 2023.

\bibitem{kadhe2014weakly}
S.~Kadhe and A.~Sprintson, ``Weakly secure regenerating codes for distributed storage,'' in \emph{2014 International Symposium on Network Coding (NetCod)}.\hskip 1em plus 0.5em minus 0.4em\relax IEEE, 2014, pp. 1--6.

\bibitem{kadhe2014weakly1}
------, ``On a weakly secure regenerating code construction for minimum storage regime,'' in \emph{2014 52nd Annual Allerton Conference on Communication, Control, and Computing (Allerton)}.\hskip 1em plus 0.5em minus 0.4em\relax IEEE, 2014, pp. 445--452.

\bibitem{paunkoska2016improved}
N.~Paunkoska, V.~Kafedziski, and N.~Marina, ``Improved perfect secrecy of distributed storage systems using interference alignment,'' in \emph{2016 8th International Congress on Ultra Modern Telecommunications and Control Systems and Workshops (ICUMT)}.\hskip 1em plus 0.5em minus 0.4em\relax IEEE, 2016, pp. 240--245.

\bibitem{paunkoska2018improving}
------, ``Improving the secrecy of distributed storage systems using interference alignment,'' in \emph{2018 14th International Wireless Communications \& Mobile Computing Conference (IWCMC)}.\hskip 1em plus 0.5em minus 0.4em\relax IEEE, 2018, pp. 261--266.

\bibitem{bian2019optimal}
J.~Bian, S.~Luo, Z.~Li, and Y.~Yang, ``Optimal weakly secure minimum storage regenerating codes scheme,'' \emph{IEEE Access}, vol.~7, pp. 151\,120--151\,130, 2019.

\bibitem{cohen2021network}
A.~Cohen, R.~G.~L. D’Oliveira, S.~Salamatian, and M.~M{\'e}dard, ``Network coding-based post-quantum cryptography,'' \emph{IEEE journal on selected areas in information theory}, vol.~2, no.~1, pp. 49--64, 2021.

\bibitem{cohen2022partial}
A.~Cohen, R.~G.~L. D’Oliveira, K.~R. Duffy, and M.~M{\'e}dard, ``Partial encryption after encoding for security and reliability in data systems,'' in \emph{2022 IEEE Int. Symp. on Inf. Theory (ISIT)}.\hskip 1em plus 0.5em minus 0.4em\relax IEEE, 2022, pp. 1779--1784.

\bibitem{lindell2017simulate}
Y.~Lindell, ``How to simulate it--a tutorial on the simulation proof technique,'' \emph{Tutorials on the Foundations of Cryptography: Dedicated to Oded Goldreich}, pp. 277--346, 2017.

\bibitem{biham1994new}
E.~Biham, ``New types of cryptanalytic attacks using related keys,'' \emph{Journal of Cryptology}, vol.~7, pp. 229--246, 1994.

\bibitem{biryukov2009related}
A.~Biryukov and D.~Khovratovich, ``Related-key cryptanalysis of the full aes-192 and aes-256,'' in \emph{Advances in Cryptology--ASIACRYPT 2009: 15th International Conference on the Theory and Application of Cryptology and Information Security, Tokyo, Japan, December 6-10, 2009. Proceedings 15}.\hskip 1em plus 0.5em minus 0.4em\relax Springer, 2009, pp. 1--18.

\bibitem{biryukov2009distinguisher}
A.~Biryukov, D.~Khovratovich, and I.~Nikoli{\'c}, ``Distinguisher and related-key attack on the full {AES}-256,'' in \emph{Annual International Cryptology Conference}.\hskip 1em plus 0.5em minus 0.4em\relax Springer, 2009, pp. 231--249.

\bibitem{bellare2003theoretical}
M.~Bellare and T.~Kohno, ``A theoretical treatment of related-key attacks: Rka-prps, rka-prfs, and applications,'' in \emph{Int. Conf. on the Theory and App. of Crypt. Techniques}.\hskip 1em plus 0.5em minus 0.4em\relax Springer, 2003, pp. 491--506.

\bibitem{biham2005related}
E.~Biham, O.~Dunkelman, and N.~Keller, ``Related-key boomerang and rectangle attacks,'' in \emph{Annual Int. Conf. on the Theory and App. of Crypt. Techniques}.\hskip 1em plus 0.5em minus 0.4em\relax Springer, 2005, pp. 507--525.

\bibitem{roetteler2015note}
M.~Roetteler and R.~Steinwandt, ``A note on quantum related-key attacks,'' \emph{Inf. Proc. Letters}, vol. 115, no.~1, pp. 40--44, 2015.

\end{thebibliography}

\end{document}